\documentclass[letterpaper]{article}
\usepackage{cite}
\usepackage{amsmath,amsfonts}
\usepackage{algorithm}
\usepackage{algorithmic}
\usepackage{array}
\usepackage[caption=false,font=normal]{subfig}
\usepackage{textcomp}
\usepackage{stfloats}
\usepackage{url}
\usepackage{verbatim}
\usepackage{graphicx}
\def\BibTeX{{\rm B\kern-.05em{\sc i\kern-.025em b}\kern-.08em
    T\kern-.1667em\lower.7ex\hbox{E}\kern-.125emX}}
\usepackage{balance}

\usepackage[rgb]{xcolor}
\usepackage{caption}
\usepackage{array}
\usepackage{enumitem}
\usepackage{hyperref}

\newcolumntype{L}[1]{>{\raggedright\arraybackslash}p{#1}}

\begin{document}

\title{Periplus: A Resilient In-band SDN Control Plane via Embedded Forwarding Graphs}

\author{E. M. Castro Barbero\textsuperscript{\rm 1}, P. de las Heras Quirós\textsuperscript{\rm 1},\\ F. J. Simó Reigadas\textsuperscript{\rm 1}
\\
\textsuperscript{\rm 1}Universidad Rey Juan Carlos}

\date{July 2026}

\maketitle

\begin{abstract}
Many resource-constrained, wide-area telecommunications deployments
could benefit from an in-band SDN control plane, but several
challenges stand in the way.  This paper presents
Periplus, an in-band SDN control plane designed to address four
challenges that this approach presents in such contexts: automatic
bootstrapping, source-based routing, fast failure recovery, and
multi-controller coordination. The
first three are developed in detail, while multi-controller
coordination is addressed in a separate paper.
 
For bootstrap, Periplus avoids network-wide flooding: when a new
switch joins, the controller installs flow rules in only two
switches. For routing and failure recovery, Periplus encodes a primary path
and per-hop alternatives in a graph structure encapsulated between L2
and L3 headers; switches forward along the primary path and, upon
detecting a failure, fall over locally to the encoded alternative
without controller involvement.
 
We evaluate a Ryu-based implementation of Periplus in Mininet across
multiple topologies. Periplus runs on stock Open vSwitch (OVS),
relying only on its built-in Nicira extensions for Network Service
Header (NSH) encapsulation. The
evaluation shows sub-50\,ms failure recovery, scalable bootstrap
across topologies of varying size and diameter, and a per-switch
flow-table footprint that is independent of network size and grows
only at switches where the controller encodes multiple alternatives.

\end{abstract}

\section{Introduction}
\label{sec:introduction}

From a conceptual standpoint, telecommunications
network architectures can incorporate mechanisms designed to ensure
reliable operation under arbitrary conditions. However, in practice,
technology tends to develop, optimize, and consolidate with specific
characteristics adapted to the demands of dominant paradigms, such as
densely populated urban areas or data centers. In these environments,
high network density and tangible economic benefits make solutions
viable that are unfeasible in more marginal scenarios. Rural
communication networks are a prime example where conventional
solutions fall short. In these deployments, low user density, reduced
profit margins, geographically dispersed systems, and limited wired
infrastructure—exacerbated by topographical and meteorological
irregularities that hinder the predictability of wireless broadband
channels—significantly impede the implementation of robust, resilient,
and highly adaptable network solutions capable of supporting Quality
of Service (QoS) with minimal specialized human intervention.

The software-defined networking (SDN)
paradigm~\cite{kreutz2014software} is an inspiring starting point for
use cases less covered by dominant technology, as it enables
centralized network control where many desired features are feasible. However, the major
development of SDN has occurred in data centers and, in general, in
networks that are primarily wired and have highly controlled and
concentrated infrastructures. Therefore, its application to more
marginal use cases as mentioned above presents a number of challenges
that must be addressed by any viable technical solution.

In most networking architectures, network functions are conceptually
divided into two planes: the control plane and the data plane. The
control plane is responsible for making decisions on how network
traffic should be managed, while the data plane forwards traffic
according to those decisions. In the Internet, although control and
data planes are conceptually separated, they have traditionally been
bundled together within each networking device. SDN introduces a
fundamental shift by proposing the physical separation of control and
data planes into distinct devices. Control plane functions are
provided by software processes running on computing nodes known as
controllers, while data plane functions are implemented by generic
programmable networking devices. Protocols such as
OpenFlow~\cite{mckeown2008openflow, specification2015open} and
P4Runtime~\cite{p4runtime2025} are used in SDN architectures to enable
communication between controllers and data plane devices, with
OpenFlow being the predominant protocol in current deployments.

SDN controllers require a control network to exchange control traffic
with data plane devices and to coordinate with other controllers. Two
primary strategies have been proposed for implementing the control
network: out-of-band and in-band~\cite{jalili2017comprehensive}. In
addition, hybrid approaches combining elements of both strategies have
also been suggested~\cite{amin2018hybrid}.

In the out-of-band strategy, the control network used to exchange
control traffic is physically separated from the network used by the
data plane to forward packets. Data plane devices are typically
connected to the controller through a dedicated port, providing direct
and isolated connectivity for control traffic.

In the in-band strategy, control traffic is forwarded by the same
networking devices that carry data plane traffic. Since data plane
devices have no dedicated channel to reach the controller, additional
protocols and mechanisms are required to bootstrap the network,
compute routing paths, achieve fast failure recovery, and coordinate
among multiple controllers.
  
Prior research highlights the critical role of the in-band strategy in
scenarios where the deployment of a separate physical infrastructure
for control traffic is impractical or economically unjustified. Such
scenarios commonly arise when SDN is deployed over wide-area networks,
including rural backbone networks~\cite{lorenti2024rural}, LEO
satellite constellations~\cite{ningyuan2021sdn}, or
industrial environments~\cite{henneke2016analysis,
  heise2017self}. Moreover, even in environments that do not exhibit
such constraints, evidence suggests that an in-band control approach
may still be preferable to achieve enhanced network flexibility,
increased programmability, lower operational and deployment costs, and
improved overall system reliability~\cite{surveyUAH}. However,
extensive research has highlighted four significant design challenges
inherent to this approach, addressing them with varying degrees of
maturity: automatic bootstrapping, efficient routing, fast failure
recovery, and multi-controller coordination.

For automatic bootstrapping, several proposals build a spanning tree
from each controller over its managed switches:
Medieval~\cite{schiff2016ground}, its self-stabilizing successor
Renaissance~\cite{canini2022renaissance}, and Sakic et
al.~\cite{sakic2020automated}, who include an RSTP-based hybrid
scheme. Wong and Lee~\cite{wong2023design} address the same challenge
in P4 networks.

For routing, Sharma et al.~\cite{sharma2016band} establish controller
sessions over the shortest path through already-connected switches.
Amaru~\cite{lopez2019amaru} introduces hierarchical labels that encode
multiple paths to each switch, supporting source-based routing without
requiring the controller to recompute paths after each topology
change.

For fast failure recovery, certain networks require recovery times
below 50\,ms \cite{RFC5654}; in in-band control planes,
restoration~\cite{vasseur2004network} is unsuitable because it
requires controller communication over the failed network. Sharma et
al.~\cite{sharma2013automatic, sharma2013fast} demonstrated this and
proposed a protection mechanism combining BFD~\cite{RFC5880} for
failure detection with OpenFlow fast-failover groups. Sakic et
al.~\cite{sakic2020automated} duplicate control traffic over multiple
paths instead. Amaru~\cite{lopez2019amaru} encodes multiple paths per
switch but loses controller-to-switch connectivity if a failure
affects the path the controller knows.

Despite these and other contributions, as far as we know, no existing
proposal addresses all four challenges simultaneously: bootstrapping
that scales without network-wide flooding or growing per-switch state;
source-based routing that retains controller reachability after a
failure; sub-50\,ms recovery without preinstalling backup paths on
every switch; and multi-controller coordination whose per-switch state
is independent of the number of controllers. This paper introduces
Periplus, a comprehensive SDN solution that addresses the four
challenges, and compares it against current state-of-the-art
alternatives.

% Bootstrap in Periplus
The Periplus in-band control plane introduced in this paper addresses
all the aforementioned limitations. The contributions related to the
first three limitations are developed in this paper:
\begin{itemize}
  \item A bootstrapping mechanism that installs flow rules in only two
    switches when a new switch joins, without network-wide
    flooding. For bootstrapping, Periplus leverages Proxy ARP,
    Anycast, and OpenFlow Packet-In/Packet-Out messages. An unmanaged
    switch broadcasts an ARP request for the anycast controller
    address; its nearest managed neighbor (the anchor) replies on
    behalf of the controller, allowing the switch to learn its
    upstream port. The controller then identifies the new switch’s
    attachment point from the source MAC address of the TCP SYN
    segment and installs the routing flows needed to place the switch
    in the managed state.
  \item A source-based routing mechanism with minimal per-switch
    state. Periplus uses the Slick Packets
    approach~\cite{nguyen2011slick}, embedding a forwarding graph
    directly into packet headers between the L2 and L3 layers using
    NSH encapsulation~\cite{rfc8300}. The graph encodes a primary path
    and per-hop alternative paths as ordered lists of output ports,
    relying solely on standard OpenFlow mechanisms including learning
    flows and register operations~\cite{Nicira}. Only the root switch
    co-located with the controller maintains routing state for all
    switches in its domain; every other switch stores only the path to
    its own controller.
  \item A sub-50\,ms failure recovery mechanism that reroutes traffic
    locally without controller involvement. Periplus combines two
    detection mechanisms: Input Traffic Detection (ITD), based on
    OpenFlow Echo Request/Reply messages and learning flows, and the
    Bidirectional Forwarding Detection (BFD) protocol~\cite{RFC5880},
    used in conjunction with OpenFlow select group
    tables~\cite{specification2015open}. Upon detecting a failure, the
    affected switch immediately selects an alternative path from the
    graph encoded in the packet, meeting carrier-grade recovery
    requirements~\cite{vasseur2004network}.
\end{itemize}

A fourth contribution, a multi-controller coordination mechanism
whose per-controller forwarding state is confined to border switches,
is developed in detail in a separate paper, which also reports how
partitioning a network across multiple controllers scales Periplus to
dozens of switches.

Periplus is implemented in Python using the Ryu SDN
framework~\cite{ryuDoc, ryuBook} and evaluated in Mininet across
multiple topologies, demonstrating scalable bootstrap and
sub-50\,ms failure recovery.

The remainder of this paper is organized as follows.
Section~\ref{sec:related-work} reviews related work on in-band SDN
control planes. Section~\ref{sec:sdn-control-plane} presents the
Periplus design, covering the bootstrap process, the graph-based
forwarding mechanism, the failure detection and recovery strategies,
and the link discovery protocol. Section~\ref{sec:evaluation} presents an
experimental evaluation in Mininet. Finally, Section~\ref{sec:conclusions}
concludes the paper.

\section{Related Work}
\label{sec:related-work}

\noindent This section reviews the main SDN solutions proposed for implementing
in-band control planes, with particular emphasis on the three
challenges addressed in this paper: bootstrapping, routing, and
failure recovery. We analyze the design
choices and inherent limitations of existing proposals, and contrast
them with the architecture and capabilities of Periplus.

\textbf{Bootstrapping.} Schiff et al. propose Medieval, a
plug-and-play distributed control plane that supports automatic
topology discovery and management relying solely on
OpenFlow~\cite{schiff2016ground}. The system bootstraps the control
plane by incrementally constructing a spanning tree over the switches
it manages, ensuring loop-free forwarding of control traffic.

Renaissance~\cite{canini2022renaissance}, an evolution of Medieval,
introduces a self-stabilizing algorithm in which the controller
maintains a single spanning tree covering the entire control
network. Failure handling relies on a slow restoration mechanism: when
a switch loses connectivity to the controller, it must reconnect to
resume operation. In both Medieval and Renaissance, the addition of
a new switch requires all intermediate switches on the path to the
controller to install new flow rules.

Renaissance progresses in rounds: when the topology changes, affected
flow rules are removed and reinstalled across all switches. Medieval
takes the opposite approach to discovery: switches use anycast
addresses, ARP traffic is flooded throughout the network, and TCP
traffic is flooded until the OpenFlow session is established. On failure, downstream switches in Medieval
lose their state and reboot, attempting to reconnect; eventually,
new spanning trees that include them are rebuilt.

Sakic et al.~\cite{sakic2020automated} present an in-band
control-plane implementation that incrementally builds a spanning tree
rooted at the controller to ensure loop-free forwarding of control
traffic.  The authors propose several variants: one that initially
relies on the Rapid Spanning Tree Protocol (RSTP), referred to as the
Hybrid Switch (HSW) scheme, and another that operates without RSTP,
known as the Hop-by-Hop (HHC) scheme. For failure handling, their
approach duplicates control-plane traffic along multiple paths,
relying on TCP at the destination to discard redundant segments.
While this strategy reduces OpenFlow table occupancy, it increases
bandwidth consumption (potentially by more than a factor of two if
$k$-resilient paths are required) and imposes additional processing
overhead on switches and the controller, which must continually
discard duplicate TCP segments.

Su et al.~\cite{su2017fasic} propose FASIC, which builds an adaptively
reconfigured spanning tree for in-band control bootstrapping and
pre-installs backup paths on switches to achieve fast failure
recovery.  Li et al.~\cite{li2021one} propose One-Pass IBAB, a
bootstrapping scheme that allows switches to begin configuration as
soon as one neighbor has established an uplink path, without waiting
for upstream switches to complete their full configuration. This
overlapping execution yields a significant reduction in bootstrapping
time compared to level-by-level approaches. However, the downlink path from the controller to each switch relies
on MAC learning in intermediate switches: as a switch's IBAB Reply
traverses the uplink, each intermediate switch records the source MAC
address and the incoming port, accumulating one forwarding entry per
downstream switch it serves.

Alvarez-Horcajo et al.~\cite{alvarez2022iehddp} propose ieHDDP, which
simultaneously performs topology discovery and in-band control channel
establishment in hybrid SDN environments comprising both SDN and
non-SDN devices, including wireless links. During a
controller-initiated exploration phase, each traversed device learns
the port through which the discovery message arrived and uses it to
establish the in-band channel back to the controller. In Periplus, by
contrast, bootstrapping is switch-initiated: an unmanaged switch
issues a Proxy ARP request for the anycast controller address and its
nearest managed neighbor replies on its behalf, so no flooding is
required and flow rules are installed in only two switches.

\textbf{Routing.} Lopez-Pajares et al. present
Amaru~\cite{lopez2019amaru}, a system in which the controller actively
explores all possible paths to each switch to incrementally generate
multiple hierarchical labels per switch, each encoding a distinct path
to the switch.  Unlike Medieval, Renaissance, and the schemes by Sakic
et al., which require sequential switch attachment starting from those
closest to the controller, Amaru significantly reduces bootstrap time
when the entire network initializes simultaneously, though this
benefit is less pronounced when integrating a single new switch into
an operational network. Amaru encodes routing information directly in
hierarchical labels, which are stored in switches and inserted into
packet headers; forwarding follows the path encoded in each
packet. Controller-to-switch communication requires the controller to
learn at least one label per switch, for instance via ARP.  Failures
in Amaru require removing the affected label at each neighbor and
propagating update events downstream so that switches can adjust their
labels. During this process, traffic from downstream switches may be
lost. Although switches store multiple labels (and therefore multiple
paths to the controller), the controller maintains only one label per
switch. Consequently, if a failure affects that path, the controller
loses connectivity to those switches until it receives updated labels,
leading to temporary loss of controller-to-switch traffic.

IS-C~\cite{goltsmann2017towards} proposes minimum-depth spanning trees
with prefix-based labeling to minimise the number of hops between
switches and the controller. Related schemes based on
distance-dependent labels have also been
explored~\cite{khakhalin2017reliable}.  Gorkemli et
al.~\cite{gorkemli2018dynamic} propose dynamic rerouting of control
traffic in response to changing network conditions.  Holzmann et
al.~\cite{holzmann2019izzy} propose a distributed protocol combining
spanning trees with topology-dependent locators, achieving sub-100\,ms
recovery in WAN topologies.

SlickFlow~\cite{ramos2013slickflow} applies Slick Packets to
data-plane forwarding in datacenters but does not address in-band
control or fast failure detection, and requires modifications to the
OpenFlow switch implementation.

General-purpose SDN link discovery protocols have also been proposed
to map network topology. OFDPv1 has the controller send a Packet-Out
containing a Link Layer Discovery Protocol (LLDP) message to each
port on every switch; on receipt, each switch generates a Packet-In
that reports the discovered link. OFDPv2~\cite{OFDPv2} reduces the
Packet-Out count by sending a single Packet-Out per switch and
having the switch forward the LLDP through all its ports; the LLDP
and Packet-In counts stay unchanged. The Tree Exploration Discovery
Protocol (TEDP)~\cite{TEDP} and the enhanced Topology Discovery
Protocol (eTDP)~\cite{eTDP} additionally provide latency-aware path
information but require a similar or even greater number of packets.
Section~\ref{sec:c-adv} presents a quantitative comparison with the
C-Adv protocol introduced in this paper.

\textbf{Failure recovery.} Beheshti and Zhang~\cite{beheshti2012fast}
address fast failover for SDN control traffic using OpenFlow
fast-failover groups, demonstrating sub-second recovery for control
channel failures. Sharma et al. focus on achieving fast failure
recovery. Their results show that restoration-based strategies are
unsuitable when recovery times below 50 ms are required in OpenFlow
networks~\cite{sharma2013automatic, sharma2013fast,
  sharma2016band}. In these studies, the authors evaluate a protection
mechanism in which the controller proactively computes backup paths
for control-plane flows and installs them in the switches. When a
switch locally detects a link or node failure using BFD, it
immediately reroutes traffic through the preinstalled backup paths by
leveraging the OpenFlow fast-failover group mechanism.  However, their
approach does not address the automatic bootstrap of the control
network, relying instead on STP and switch-level DHCP
modifications. Su et al.~\cite{su2017fasic} similarly pre-install
backup paths to achieve fast recovery, combining this with an
adaptively reconfigured spanning tree for bootstrapping.

Table~\ref{tab:related-work-comparison} summarises the position of
each proposal across the three challenges. The table includes works
that make a primary and direct contribution to at least one of the
three challenges in an in-band SDN control plane.

\begin{table*}[hbt]
\caption{Comparison of in-band control plane proposals across the three key challenges.}
\label{tab:related-work-comparison}
\scriptsize
\centering
\setlength{\tabcolsep}{4pt}
\begin{tabular}{|L{1.5cm}|L{3.5cm}|L{2.3cm}|L{3.5cm}|}
\hline
\textbf{System} & \textbf{Bootstrapping} & \textbf{Routing} & \textbf{Failure Recovery} \\
\hline
Medieval / Renaissance \cite{schiff2016ground,canini2022renaissance} &
  Spanning tree; network-wide ARP flooding; per-switch state grows with network size &
  Destination-based &
  Restoration (slow; sub-50\,ms not achieved) \\
\hline
Sakic et~al. \cite{sakic2020automated} &
  Spanning tree (RSTP/HHC); per-switch state grows with network size &
  Destination-based &
  Traffic duplication along multiple paths \\
\hline
FASIC \cite{su2017fasic} &
  Adaptively reconfigured spanning tree &
  Destination-based &
  Protection via preinstalled backup paths \\
\hline
Wong and Lee \cite{wong2023design} &
  Automatic bootstrapping (P4 networks) &
  Destination-based &
  Protection via pre-planned recovery paths (sub-50\,ms) \\
\hline
Sharma et al. \cite{sharma2013automatic,sharma2013fast,sharma2016band} &
  Not addressed; relies on STP + DHCP pre-configuration &
  Destination-based &
  Protection via fast-failover groups (sub-50\,ms); backup paths preinstalled on all switches \\
\hline
Amaru \cite{lopez2019amaru} &
  Hierarchical label exploration (no flooding) &
  Source-based (hierarchical labels) &
  Label update propagation; temporary loss of controller-to-switch reachability possible \\
\hline
\textbf{Periplus} &
  \textbf{Proxy ARP + Anycast; flow rules in 2 switches only} &
  \textbf{Source-based (graph in header); minimal per-switch state} &
  \textbf{Protection (sub-50\,ms); local preinstallation; controller reachability maintained} \\
\hline
\end{tabular}
\end{table*}

The works reviewed above address the three challenges with varying
degrees of maturity. Bootstrapping proposals achieve plug-and-play
operation at the cost of network-wide flooding or per-switch state
that grows with network size. Source-based routing approaches such as
Amaru reduce forwarding state but cannot guarantee controller
reachability after a failure. Only a subset of proposals achieve
sub-50\,ms failure recovery, and none do so without preinstalling
backup paths on all switches. The proposals most directly comparable
to Periplus are
Amaru, which shares the source-based routing approach but risks
temporary loss of controller reachability after failures; Sharma et
al., which achieves sub-50\,ms recovery but relies on STP for
bootstrap; and Medieval/Renaissance and Sakic et al., which address
bootstrapping but impose forwarding state proportional to network
size on every switch. Periplus addresses all three challenges
simultaneously, as described in the following sections.

\section{Periplus SDN Control Plane}
\label{sec:sdn-control-plane}

\noindent This section presents the design of Periplus, which rests on three
core design choices (graph-based source routing, anchor-relayed
bootstrap, and packet-encoded failure recovery), elaborated in the
contributions of Section~\ref{sec:introduction}.

Throughout this paper, we assume that node communication interfaces
---wired or wireless---are based on IEEE 802 link technologies, so
that each interface is identified by a unique 48-bit MAC address. We
further assume that links are bidirectional: any port on which a node
receives traffic can also forward traffic over the same link in the
opposite direction.

Switches are denoted $s_i$ throughout, with $c_0$ reserved for the
OVS switch co-located with the controller (the root switch).

Controller Advertisement (C-Adv) messages serve as a cross-cutting
discovery mechanism throughout the system. They assist switch
attachment to the controller and drive the link discovery protocol
that reveals redundant links used as alternative paths; they also
underpin multi-controller coordination, which will be addressed in a
separate paper.

Section~\ref{sec:bootstrap} describes the bootstrap process;
Section~\ref{sec:forwarding-graphs} presents the graph-based
forwarding mechanism; Sections~\ref{sec:monitoringports}
and~\ref{sec:c-adv} cover failure detection and link discovery.

A prototype of Periplus has been implemented in Python using the Ryu
framework~\cite{ryuDoc, ryuBook}. The implementation leverages the
Nicira Extension Action Structures for OpenFlow~\cite{Nicira} for
NSH encapsulation and decapsulation, register manipulation, and
dynamic learning flows in OVS.

\subsection{Bootstrap}
\label{sec:bootstrap}

\noindent Periplus bootstraps the control plane incrementally: switches attach
to the controller one by one, each relying on a neighboring managed
switch to relay their connection requests.

The controller runs on a node that also hosts an OVS
instance, configured with an anycast address known by all SDN
switches. Initially, only the root switch $c0$ is managed; its
OpenFlow session is established directly through the local TCP/IP
stack, after which the controller installs the flow rules that place
$c0$ in the managed state.

Unmanaged switches rely on the cooperation of a neighboring managed
switch, referred to as the anchor, which encapsulates their packets
into Packet-In messages destined for the controller.

\begin{figure*}[t]
\centering
\includegraphics[width=\linewidth]{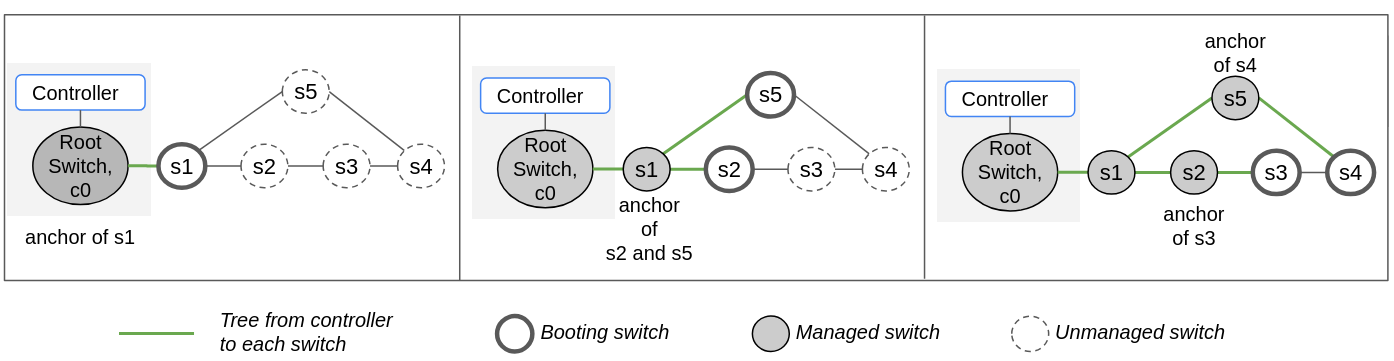}
\caption{Incremental bootstrap. Initially only the root switch $c0$ is
managed and $s1$ boots using $c0$ as its anchor (left). Once $s1$ is
managed, $s2$ and $s5$ can proceed (middle). Once $s2$ and $s5$ are
managed, $s3$ and $s4$ can proceed (right).}
\label{fig:anchors}
\end{figure*}

Fig.~\ref{fig:anchors} illustrates the process: each switch can attach
only once a managed neighbor is available to serve as its anchor, so
bootstrapping propagates wave by wave across the topology until all
switches are managed.

Each switch attachment follows two phases: first, ARP discovery to
identify the upstream port toward the controller; second, TCP
connection establishment and OpenFlow handshake, after which the
controller installs the routing flows that place the switch in the
managed state.

\subsubsection{ARP}

\noindent Initially, only the root switch operates under controller
management. The unmanaged switches are preconfigured with a minimal
set of flow rules that provide basic functionality: specifically, they
broadcast ARP requests on all their interfaces to discover the
controller’s MAC address (see Table~\ref{tab:broadcastARP}).

\setlength{\tabcolsep}{3pt}
\begin{table}[h]
\centering
\caption{Broadcast ARP request.}
\label{tab:broadcastARP}
  \scriptsize
  \begin{tabular}{|l|c|l|l|}
    \hline
	  \textbf{Name} &\textbf{Table ID} & \textbf{Match} & \textbf{Action}\\
\hline
	  arp\_bcast & 2& arp, arp\_op=1, in\_port=LOCAL & ALL\\
\hline
\end{tabular}
\end{table}

Neighboring switches reply to the ARP request only if they are managed
switches. These switches are configured with a specific flow rule that
enables them to act as proxy ARP agents for ARP messages targeting the
controller’s MAC address. As shown in Table~\ref{tab:proxyARP},
where \texttt{C\_IP} denotes the anycast controller’s IP address and
\texttt{C\_MAC} denotes the controller’s MAC address, this flow
rule extracts the fields from an incoming ARP request,
rewrites them to generate an ARP reply, and forwards it to the input
port \texttt{IN\_PORT}. Every managed switch answers with the same
\texttt{C\_MAC}, so the requesting unmanaged switch learns a single
MAC address for the controller regardless of which neighbour
proxied the reply — this is what makes anycast addressing of the
controller work.

\begin{table}[h]
\centering
\caption{Proxy ARP in managed switches.}
\label{tab:proxyARP}
  \scriptsize
  \begin{tabular}{|l|c|l|l|}
    \hline
	  \textbf{Name} & \textbf{Table ID} & \textbf{Match} & \textbf{Action}\\
\hline
 proxy\_arp & 2 &
arp,    & decap(packet\_type(ns=0,type=0)),\\
& &  arp\_tpa=C\_IP, & push: NXM\_OF\_ARP\_SHA[],\\
&&  arp\_op=1, & push: NXM\_OF\_ARP\_SPA[],\\
	  &&& push: NXM\_OF\_ARP\_SHA[],\\
	  &&& set\_field: C\_IP$\rightarrow$ arp\_spa,\\
	  &&& set\_field: C\_MAC$\rightarrow$ arp\_sha,\\
	  &&& pop: NXM\_NX\_ARP\_THA[],\\
	  &&& pop: NXM\_OF\_ARP\_TPA[],\\
	  &&& set\_field: arp\_op=0x2,\\
	  &&& encap(ethernet),\\
	  &&& pop:NXM\_OF\_ETH\_DST[], \\
	  &&& set\_field: C\_MAC $\rightarrow$ eth\_src, \\
	  &&& output:IN\_PORT\\
\hline
\end{tabular}
\end{table}

As a result, any unmanaged switch with at least one managed neighbor
can receive an ARP reply from that neighbor, allowing it to learn the
controller’s MAC address and identify the corresponding upstream
interface.

\subsubsection{TCP connection establishment}

\begin{figure}[!h]
    \centering
	\includegraphics[width=0.6\linewidth]{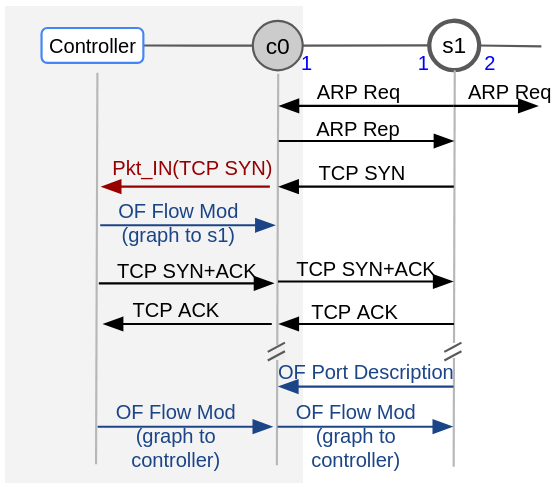}
    \caption{Message exchange during the bootstrap of switch $s1$ via anchor switch $c0$ (TCP three-way handshake plus OpenFlow handshake).}
	\label{fig:TCPs1-c0}
\end{figure}

\noindent Once the upstream port is identified during the ARP phase, the switch
initiates a TCP connection to establish an OpenFlow session with the
controller. Fig.~\ref{fig:TCPs1-c0} illustrates the full message
exchange between a new unmanaged switch \texttt{s1} and the
controller, with the assistance of the anchor switch \texttt{c0} that
has previously established both a TCP connection and an OpenFlow
session with the controller.

Upon receiving the SYN segment from an unmanaged switch $s1$, the
anchor switch $c0$ generates a Packet-In message and relays it to
the controller, which infers the presence of a previously unknown
link between $c0$ and $s1$. Leveraging this information, the
controller incrementally reconstructs the network topology and
computes the path required to reach the newly discovered switch.

The controller then installs a new flow entry in the root switch $c0$
that applies to all packets addressed to the newly detected switch
$s1$. This entry associates the computed path with all traffic
destined for $s1$, enabling all switches in the path to select the
correct output interface.

After the TCP connection is established, the controller and the switch
exchange the initial OpenFlow session messages, including the
\texttt{OFPMP\_PORT\_DES\-CRIPTION} reply, which reports the per-port
hardware addresses of the new switch. The controller determines the
upstream port used by the switch by matching the source MAC
address of the received TCP SYN segment against those hardware
addresses; in the example, this resolves to port~1 of $s1$.
Finally, the controller installs a flow entry that embeds the
forwarding graph into packets originating from $s1$ and destined for
the controller.

The exchange just described presupposes that the unmanaged switch has
already identified the physical port through which to emit its first
SYN segment. This identification must be performed by the switch
itself, before any controller-installed flow exists. Periplus
addresses this by observing that the ingress port of any frame whose
origin can be unambiguously attributed to the controller side of the
network is, by construction, the upstream port toward the controller.
Two control-plane signals admit such unambiguous attribution in our
setting, and Periplus installs a learning rule for each
(Table~\ref{tab:learnedReg9}).

\begin{table}[hb]
\centering
\caption{Preconfigured learning rules: cache the upstream port toward the controller upon receipt of an ARP reply or a C-Adv.}
\label{tab:learnedReg9}
  \scriptsize
  \begin{tabular}{|l|c|l|l|}
    \hline
	  \textbf{Name} & \textbf{Table ID} & \textbf{Match} & \textbf{Action}\\
\hline
	  arp\_learn &  2 & reg9=0,arp,  & learn(table=1, hard\_timeout=10, \\
	  && arp\_tpa=SW\_IP, & \hspace{0.5cm} load:\\
	  && arp\_spa=C\_IP & \hspace{1cm}OXM\_OF\_INPORT[] $\rightarrow$ \\
	  &&& \hspace{1cm}NXM\_NX\_REG9[])\\
\hline
	  C-Adv\_learn &2 &
reg9=0, udp  & learn(table=1, hard\_timeout=10, \\
	  && tp\_src=55555, & \hspace{0.5cm} load:\\
	  && tp\_dst=55556 & \hspace{1cm}OXM\_OF\_INPORT[] $\rightarrow$ \\
	  &&& \hspace{1cm}NXM\_NX\_REG9[])\\
\hline
\end{tabular}
\end{table}

A unicast ARP reply matching \texttt{arp\_spa=C\_IP},
\texttt{arp\_tpa=SW\_IP} (rule \textit{arp\_learn}) can only have been
generated by a managed neighbor acting as proxy ARP agent for the
controller (rule \textit{proxy\_arp}, Table~\ref{tab:proxyARP}), in
response to the unmanaged switch's own broadcast ARP request. Since
only managed switches install the proxy-ARP rule, and a managed switch
by definition already sits on a path to the controller, the reply's
ingress port is necessarily controller-facing. A Controller
Advertisement, identified by the dedicated UDP port pair
\texttt{tp\_src=55555}, \texttt{tp\_dst=55556} (rule
\textit{C-Adv\_learn}), is emitted only by the controller and
re-forwarded only by managed switches outward toward their unmanaged
neighbors; any C-Adv arriving at an unmanaged switch therefore came
from a neighbor that already sits on a path to the controller, and its
ingress port is again controller-facing. The full C-Adv protocol is
described in Section~\ref{sec:c-adv}.

Either trigger is sufficient on its own, but Periplus installs both
because they cover complementary cases. The ARP rule handles the
cold-start case, when the switch boots without any cached state. The
C-Adv rule handles the warm-restart case: a switch that was previously
managed still has the controller's MAC address in its ARP cache,
issues no ARP request, and would otherwise be unable to rediscover its
upstream port; a C-Adv from a managed neighbor then becomes the only
available signal.

Both rules install the same learning action: load the ingress port
into a soft-state slot with a 10-second hard timeout. In the Open
vSwitch realization this slot is the register \texttt{reg9}, exposed
by the \texttt{learn} action; its role is purely that of a cache for
the most recently observed upstream port. The 10-second timeout is
sufficient to complete the OpenFlow handshake in the evaluated
environments, during which the controller installs higher-priority
routing flows that supersede the soft state. The timeout is
configurable: higher-latency links may require a larger value, at the
cost of a longer window during which the cached entry persists if the
connection attempt fails. If no further ARP reply or C-Adv arrives
(e.g., when the controller is stopped), the entry expires and the
switch returns to the unknown-upstream state.

\begin{figure}[ht]
\centering
	\includegraphics[width=\linewidth]{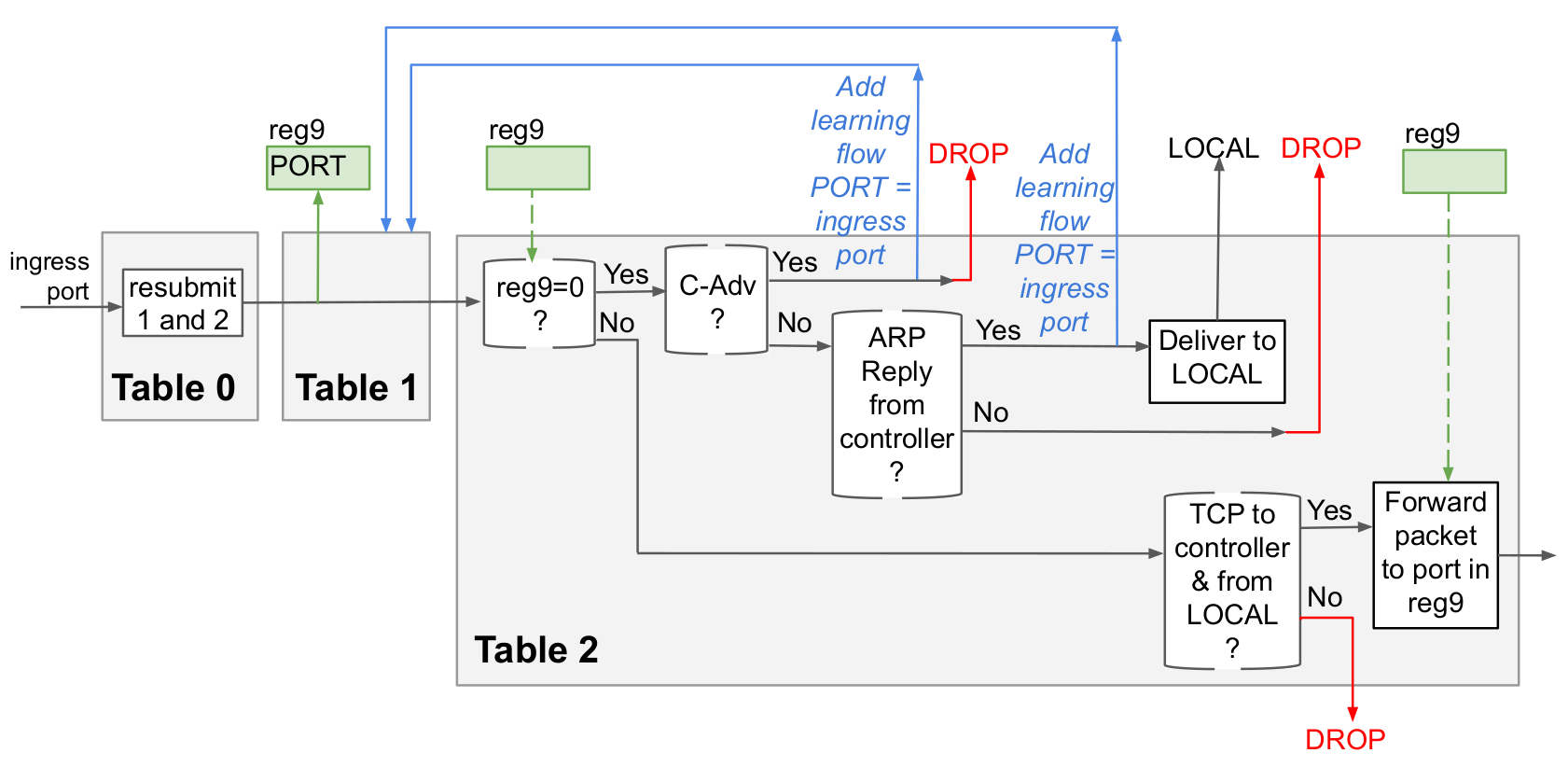}
\caption{OVS pipeline: the upstream port toward the controller, learned at table~1, is used by table~2 to forward control-plane TCP traffic.}
\label{fig:localTCPreg9}
\end{figure}

Fig.~\ref{fig:localTCPreg9} illustrates how the cached upstream port
flows through the OVS pipeline. Locally generated TCP
segments enter at table~0 and are resubmitted to tables~1 and~2. At
table~1 the learning flow installed by either trigger reloads the
cached port into the register on every packet; at table~2 the
forwarding rules of Table~\ref{tab:useReg9} consume that port to send
the segment out toward the controller.

Table~\ref{tab:useReg9} lists the rules that consume the learned
upstream port to forward the switch's own control-plane TCP traffic;
the value \texttt{ETH\_ADDR} is the MAC address of that port.
The SYN segment requires special treatment: before being emitted, its
source MAC address is rewritten to \texttt{ETH\_ADDR}, so that
the controller can later infer, from the Packet-In delivered by the
anchor, the link over which the new switch is attaching (rule
\textit{send\_syn}). Subsequent segments of the same connection are
forwarded through the same port without the rewrite (rule
\textit{send\_tcp}). If no upstream port has yet been learned, the SYN
is dropped (rule \textit{drop\_not\_ctrlr}), and the switch will retry
once the next ARP reply or C-Adv arrives.

\begin{table}[h]
\centering
        \caption{OVS rules that forward control-plane TCP segments through the learned upstream port.}
\label{tab:useReg9}
  \scriptsize
	\begin{tabular}{|l|c|l|l|}
    \hline
		\textbf{Name} & \textbf{Table ID} & \textbf{Match} & \textbf{Action}\\
\hline
		go\_tables\_1\_2& 0 & & resubmit(,1), resubmit(,2)\\
\hline
		save\_ctrlr\_port & 1 & & load:\\
		&& & \hspace{0.1cm}PORT $\rightarrow$ \\
		&& & \hspace{0.1cm}NXM\_NX\_REG9[0..15]\\
\hline
		drop\_not\_ctrlr & 2 &
tcp, in\_port=LOCAL,  & DROP  \\
		&& reg9=0, nw\_dst=C\_IP  &  \\
\hline
		send\_syn & 2 &
tcp, in\_port=LOCAL, & eth\_field:\\
		&& nw\_dst=C\_IP,  & \hspace{0.1cm} ETH\_ADDR $\rightarrow$ eth\_src,\\
		&& tcp\_flags=tcp.TCP\_SYN & output: \\
		&& reg9=PORT &  \hspace{0.1cm}NXM\_NX\_REG9[0..15] \\
\hline
		send\_tcp &2 & tcp, in\_port=LOCAL, & output:\\
		&& nw\_dst=C\_IP &  \hspace{0.1cm}NXM\_NX\_REG9[0..15] \\
 \hline

\end{tabular}
\end{table}

%%%%%%%%%%%%%%%%%%%%%%%%%%%%%%%%%%%%%%%%%%%%%%%%

\subsection{Forwarding based on graphs}
\label{sec:forwarding-graphs}

\noindent Periplus uses the Slick Packets approach~\cite{nguyen2011slick} to
route control traffic. A forwarding graph is an acyclic subgraph of
the network topology, embedded in packet headers, that encodes
multiple paths between source and destination.

The graph encodes a primary path $P$ and, at each hop along $P$, a
list of alternative paths. Let $n$ be the number of switches along
the path, indexed $i = 1, \dots, n$ in order. The primary path $P =
[port_1, \dots, port_n]$ assigns to each node $i$ an output port
$port_i$. For each node $i$, an ordered (possibly empty) list of
alternative paths $\mathcal{A}_i = [A_i^{(1)}, \dots, A_i^{(K_i)}]$ is
also included, where $K_i \geq 0$ is the number of alternative paths
the controller encodes at node $i$ (at most as many as the topology
permits). Each $A_i^{(k)}$ encodes an output-port sequence from
node $i$ to the destination. When node $i$ cannot forward through
$port_i$, it tries $A_i^{(1)}$; if the first port of $A_i^{(1)}$
is also unavailable, it falls back to $A_i^{(2)}$, and so on
through $A_i^{(K_i)}$. Each $port_i$ denotes an OpenFlow output port
number at node $i$; the controller computes the graph in terms of the
port numbers reported by each switch via OpenFlow.

Periplus encodes $P$ in a single NSH header~\cite{rfc8300} inserted
between the L2 and L3 headers, with each port entry flagged to
indicate whether one or more alternative paths exist for that hop.
Each $A_i^{(k)}$ is encoded in a separate NSH header. NSH was chosen
because OVS~\cite{OpenvSwitch} provides native support for NSH
encapsulation and decapsulation via Nicira
extensions~\cite{Nicira}, and its context header fields offer
sufficient space to encode port sequences without a custom Ethertype
or modifications to the switch firmware. As the packet progresses
along $P$, the $A_i^{(k)}$ headers for each successfully traversed
hop are discarded, since those alternatives are no longer reachable.

When the controller learns of a new switch, it computes forwarding
graphs for both traffic directions. The primary path is the shortest
path obtained from Dijkstra's algorithm on the known topology. At
each hop, up to $K_i$ alternatives are obtained by iteratively running
Dijkstra and pruning, at each iteration, the first edge of every
previously-chosen alternative, yielding source-disjoint alternatives
where the topology permits.

The per-hop count $K_i$ is a controller-side policy choice. The
controller may assign different values of $K_i$ to different hops
within the same graph, and different policies to different (source,
destination) pairs. Each additional alternative encoded at hop $i$
adds one NSH header to the packet (24 bytes per $A_i^{(k)}$) and
installs additional flow rules only at the hop-$i$ switch, leaving
every other switch unaffected. At the source, a packet carries one
header for $P$ plus $\sum_i K_i$ headers for the alternatives---in
the worst case $1 + nK$ headers ($24(1+nK)$ bytes) for a uniform
$K_i = K$ over $n$ hops---and this count diminishes hop by hop as
alternatives no longer reachable are stripped. As in Slick
Packets~\cite{nguyen2011slick}, natural policies include a uniform
$K_i$ across hops, per-destination policies that raise $K_i$ only
for high-availability destinations, and topology-aware policies
that raise $K_i$ at hops where the $K = 1$ alternative shares a
failure domain with downstream primary segments. The prototype
implements one such instance, $K_1 = 2$ and $K_i = 1$ for $i > 1$,
evaluated in Section~\ref{sec:evaluation} and motivated by the
cascading-failure pattern of Section~\ref{sec:cascadingFailure}
where the primary first hop and the first hop of $A_1^{(1)}$ may
fail in the same incident, leaving the data plane with no
preinstalled alternative.

For the controller-to-switch direction, flow rules are installed in
the root switch to embed the computed graph into every control packet
the controller sends to that switch. For the switch-to-controller
direction, flow rules are installed in the switch itself to embed
the corresponding graph into control packets destined for the
controller.

The root switch serves a distinct role: all traffic to and from the
controller passes through it. Because the controller and the root
switch share the same node, they communicate through the local TCP/IP
stack without requiring forwarding graphs. From the perspective of
OpenFlow table occupancy, the root switch maintains one flow per
managed switch to embed the correct graph into controller-originating
packets; every other switch stores only the single flow that embeds
its own path to the controller.

When a packet arrives at a switch, the forwarding logic follows
Algorithm~\ref{alg:periplus_forwarding}.

\begin{algorithm}
\caption{Periplus Adaptive Packet Forwarding}
\footnotesize
\label{alg:periplus_forwarding}
\begin{algorithmic}[1]
	\STATE Let $P = [port_1, \dots, port_n]$ be the primary path
	\STATE Let $\mathcal{A} = [\mathcal{A}_1, \dots, \mathcal{A}_n]$
        with $\mathcal{A}_1 = [A_1^{(1)}, \dots, A_1^{(K_1)}]$ the
        alternative paths at this hop
\IF{$port_1$ is active}
    \STATE $P \gets (port_2, \dots, port_n)$
	\STATE $\mathcal{A} \gets [\mathcal{A}_2, \dots, \mathcal{A}_n]$
    \STATE \textbf{forward} via $port_1$
\ELSE
    \FOR{$k = 1$ \TO $K_1$}
        \IF{$A_1^{(k)} \neq []$ \AND the first port of $A_1^{(k)}$ is active}
            \STATE $P \gets A_1^{(k)}$
            \STATE $\mathcal{A} \gets []$
            \STATE \textbf{reapply forwarding}; \textbf{return}
        \ENDIF
    \ENDFOR
    \STATE \textbf{drop} packet
\ENDIF
\end{algorithmic}
\end{algorithm}

In the $K_1 = 2$ first-hop policy evaluated in
Section~\ref{sec:evaluation}, the source switch forwards via the
primary path when $port_1$ is up and, when $port_1$ is down,
selects $A_1^{(1)}$, falling through to $A_1^{(2)}$ if the first
port of $A_1^{(1)}$ is also down, as in
Algorithm~\ref{alg:periplus_forwarding}. The latter case is the
cascading-failure pattern that motivates $K_1 \geq 2$, where a
single incident disables both $port_1$ and the first port of
$A_1^{(1)}$. Downstream switches see $K_i = 1$ and behave as in the
baseline single-alternative case.

\begin{figure*}[!t]
    \centering

    \subfloat[Graph forwarding from $s3$ to the controller.\label{fig:grafos-un-camino}]
    {
        \includegraphics[width=0.7\textwidth]{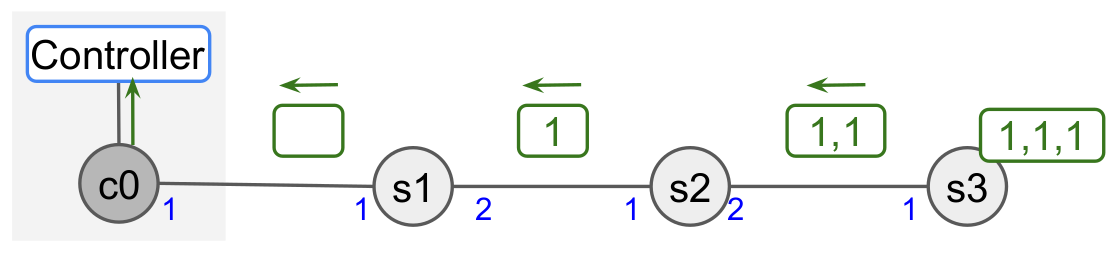}
    }\\[2ex]
    \subfloat[All ports active; $A_1$ and $A_2$ headers are discarded as the packet progresses along $P$.\label{fig:alternativo-1}]
    {
        \includegraphics[width=0.7\textwidth]{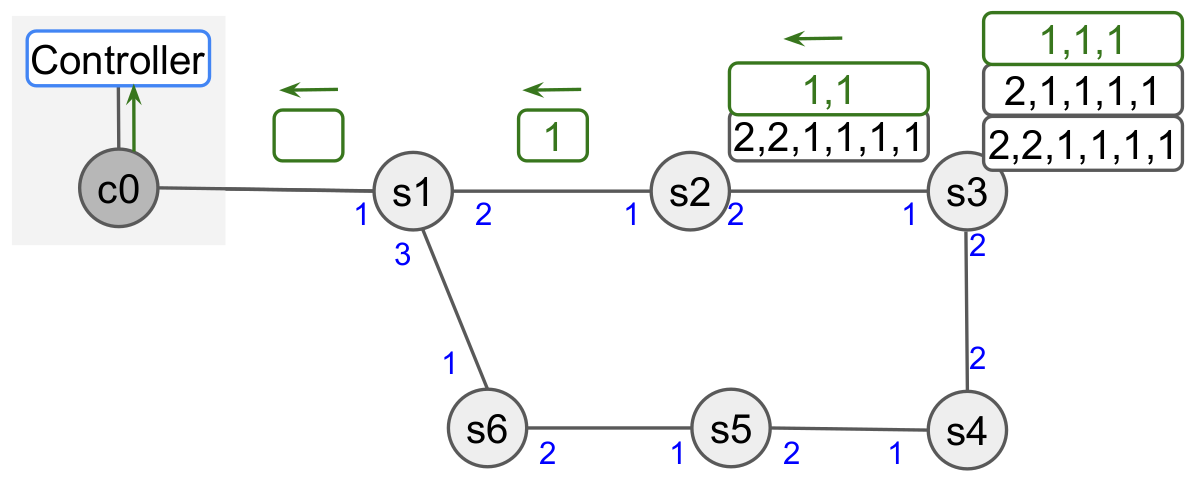}
    }\\[2ex]
    \subfloat[Port~1 of $s3$ inactive; $A_1$ is activated.\label{fig:alternativo-2}]
    {
        \includegraphics[width=0.7\textwidth]{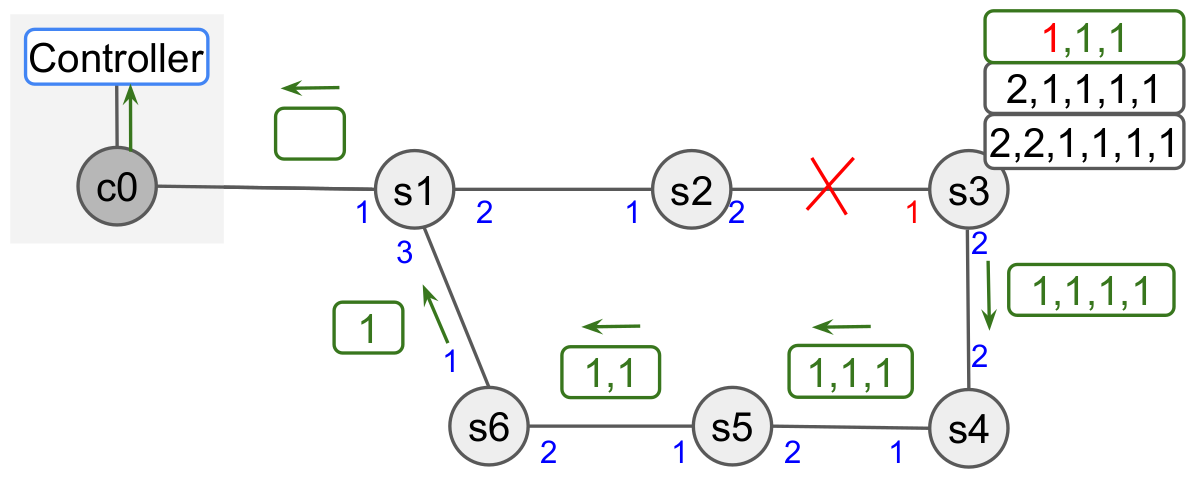}
    }\\[2ex]
    \subfloat[Port~1 of $s2$ inactive; $A_1$ discarded at $s3$, $A_2$ activated at $s2$.\label{fig:alternativo-3}]
    {
        \includegraphics[width=0.7\textwidth]{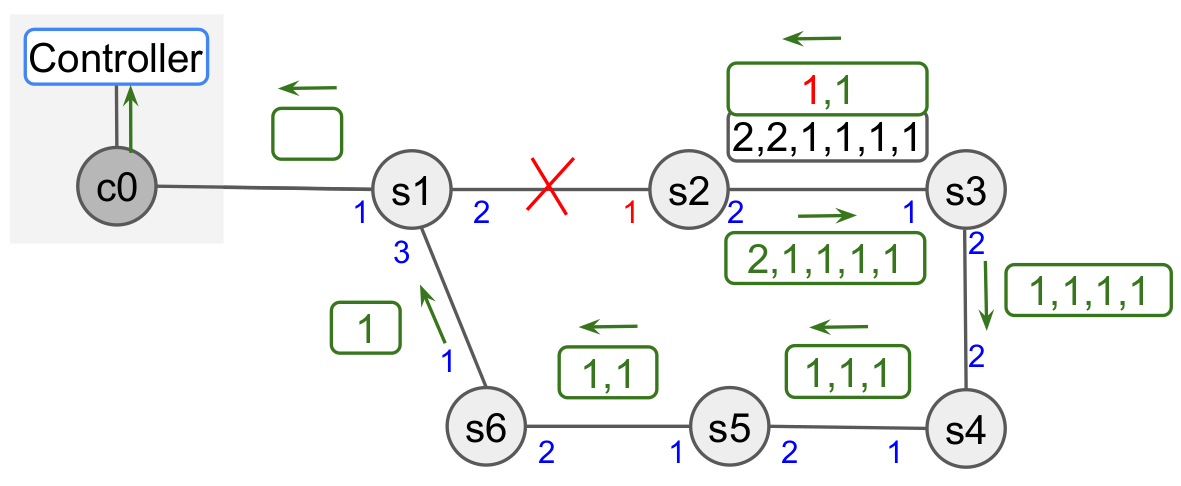}
    }

    \caption{Graph forwarding from $s3$ to the controller. Port
      numbers (in blue) are shown close to the corresponding
      links. The primary path $P$ appears in the top green box at each
      switch; alternative paths $A_1$, $A_2$, \ldots\ (one per hop of
      $P$) appear in the boxes below, and are discarded as the packet
      progresses along $P$.}
    \label{fig:grafos}
\end{figure*}

Under normal conditions, packets follow the primary path. When a
switch detects that its designated output port is inactive, it
immediately switches to the corresponding alternative path without
contacting the controller (the protection mechanism). Port liveness
is determined by the BFD or ITD
mechanisms described in Section~\ref{sec:monitoringports}. Once the
controller learns of the failure, it recomputes routes and installs
new primary paths that avoid the failure, providing a complementary
restoration mechanism. Ongoing traffic is uninterrupted during this
transition.

The following examples illustrate Algorithm~\ref{alg:periplus_forwarding}
on the topologies of Fig.~\ref{fig:grafos}. In the examples that
follow, each hop encodes at most one alternative ($K_i = 1$), so we
write $A_i$ as shorthand for $A_i^{(1)}$. The multi-alternative
first-hop case ($K_1 = 2$) is revisited in
Section~\ref{sec:cascadingFailure}.

Fig.~\ref{fig:grafos-un-camino} shows the single-path case: $s3$
encodes $P = [1, 1, 1]$ in NSH and strips its own output port before
sending; subsequent switches do the same until the packet reaches
$c0$.

Fig.~\ref{fig:alternativo-1} adds alternative paths
$A_1 = [2, 1, 1, 1, 1]$ and $A_2 = [2, 2, 1, 1, 1, 1]$ ($A_3 = []$,
no alternative from $s1$). With all ports active, the $A_i$ headers
are discarded as the packet progresses along $P$.

Fig.~\ref{fig:alternativo-2}: $s3$ detects port~1 inactive and adopts
$A_1$ as the new primary path, forwarding through $s4, s5, s6, s1$ to
$c0$.

Fig.~\ref{fig:alternativo-3}: $s3$ forwards through port~1 normally,
but $s2$ then detects its link to $s1$ inactive and adopts the
remaining alternative (originally $A_2$), routing backward through
$s3$ before reaching $c0$ via $s4, s5, s6, s1$.

%%%%%%%%%%%%%%%%%%%%%%%%%%%%%%%%%%%%%%%%%%%%%%%%
\subsection{Monitoring ports}
\label{sec:monitoringports}

\noindent Periplus provides two mechanisms to detect link failures and trigger
the selection of an alternative path by the switch:

\begin{enumerate}[label=(\roman*)]
\item Bidirectional Forwarding Detection (BFD).
\item Input Traffic Detection (ITD).
\end{enumerate}

BFD is suited to environments where the switch supports the protocol
natively and direct link monitoring between adjacent nodes is
available. ITD provides an alternative when BFD is unavailable,
inferring port liveness from the presence of incoming traffic. Both
are implemented using OVS-specific OpenFlow extensions: BFD
relies on select-group tables with \texttt{watch\_port}, while ITD
uses learning flows to dynamically track port activity.

In both mechanisms, whenever the switch receives a packet carrying an
NSH header, it extracts the output port associated with the primary
path and verifies whether this port is operational.

\subsubsection{BFD}
\label{sec:bfd}
\noindent OVS supports the BFD protocol, which identifies failures
between two forwarding nodes connected by a direct link. When a BFD
session detects a link fault, it transitions to the down
state. Periplus leverages OpenFlow group tables to monitor the
operational state of switch ports. Specifically, select-type groups
are used to determine whether the output port associated with the
primary path is operational. The group employs the
\texttt{watch\_port} option to verify the activity status of the port.

\begin{figure}[ht]
\centering
	\includegraphics[width=\linewidth]{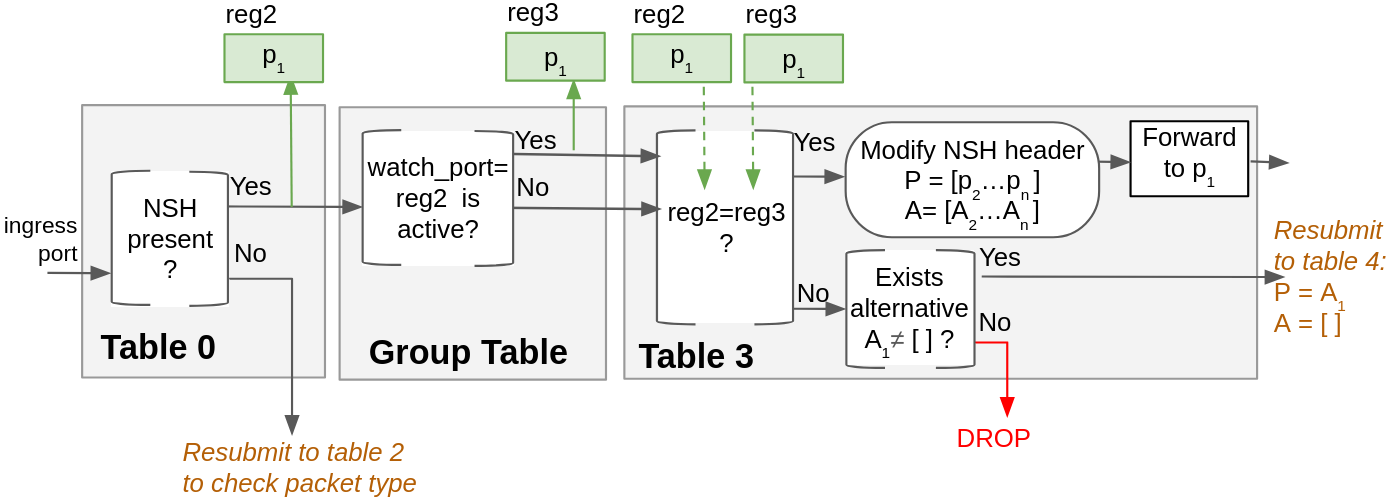}
	\caption{OVS group-table mechanism that monitors the operational state of an output port using BFD.}
\label{fig:BFD}
\end{figure}

In Fig.~\ref{fig:BFD}, the primary path for a packet being processed
specifies port $p_1$ as the output port. It is stored in $reg2$.  If
$p_1$ is active, the corresponding action stores $p_1$ also in
register $reg3$ for subsequent processing. Conversely, if $p_1$ is
inactive, the action is not executed, and $reg3$ retains the value
zero. Before the packet is forwarded, if $reg2$ and $reg3$ match, the
packet is forwarded via $p_1$; otherwise, if the alternative path
$A_1$ is available, it is selected instead.

\subsubsection{ITD}
\label{sec:itd}

\begin{figure}[ht]
\centering
	\includegraphics[width=\linewidth]{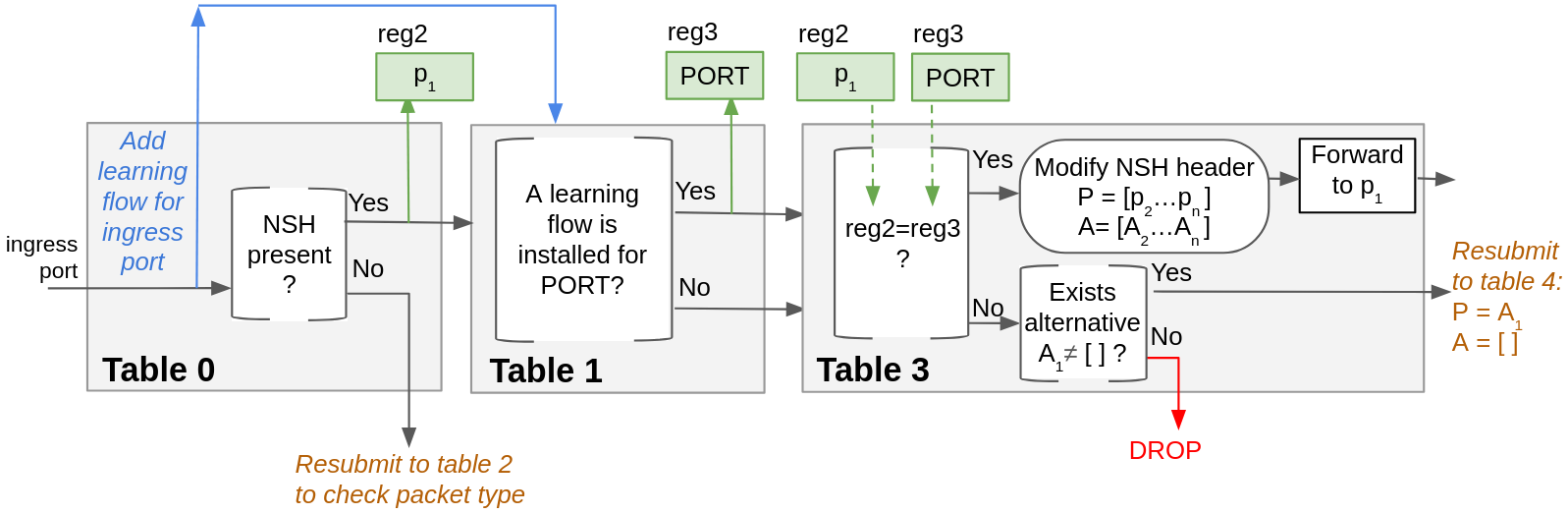}
	\caption{OVS pipeline used by the Input Traffic Detection (ITD) mechanism to track per-port liveness via learning flows.}
\label{fig:ITD}
\end{figure}

\noindent ITD infers port liveness from the presence of incoming traffic: if a
switch receives packets through a given port within a configurable
time window, that port is considered operational. OpenFlow echo
request/reply messages exchanged between the controller and each
switch provide a steady source of traffic that keeps monitored ports
alive under normal conditions.

Fig.~\ref{fig:ITD} illustrates the traversal of a packet through the
OVS pipeline when ITD is employed. Upon arrival at any port,
the packet first enters OpenFlow table ID 0, after which a learning
flow with a hard timeout of 2 seconds is installed in OpenFlow table
ID 1. The action associated with this learning flow updates $reg3$ to
indicate that the ingress port (\texttt{PORT}) for the packet is
active (see Table~\ref{tab:learningFlow}). If no subsequent packets
are received through that ingress port before the timeout expires, the
learning flow is removed. Thus, the presence of this flow in OpenFlow
table ID 1 indicates that the corresponding port is currently
operational.

Future packets whose primary-path output port (stored in $reg2$) is
\texttt{PORT} will match the corresponding learning flow installed in
OpenFlow table ID 1, causing \texttt{PORT} to be written into
$reg3$. If $reg2$ equals $reg3$, the switch determines that $p_1$ is
active. However, if the packet does not match the associated learning
flow in OpenFlow table ID 1, the value of $reg3$ remains zero. In this
case, the switch forwards the packet through the alternative path, if
one is available; otherwise, the packet is dropped.

\begin{table}[h]
\caption{Learning action in the ITD mechanism.}
\label{tab:learningFlow}
  \centering
  \scriptsize
	\begin{tabular}{|l|c|l|l|}
    \hline
		\textbf{Name} & \textbf{Table ID} & \textbf{Match} & \textbf{Action}\\
\hline
		active\_port& 0 &
	  in\_port=\scriptsize{PORT} & learn(table=1, hard\_timeout=2,\\
		& & & \hspace{0.7cm}eth\_type=0x0800,\\ 
		& & & \hspace{0.7cm}reg2=\scriptsize{PORT}, \\
		& & & \hspace{0.7cm}load:\\
		& & & \hspace{1cm}\scriptsize{OXM\_OF\_IN\_PORT[]}$\rightarrow$\\
		& & & \hspace{1cm}\scriptsize{NXM\_NX\_REG3[])},\\
		& & & resubmit(,1), \\
		& & & resubmit(,3)\\
\hline
\end{tabular}
\end{table}

Fig.~\ref{fig:ITD} also illustrates the pipeline processing for
control-plane packets. In OpenFlow table ID~0, if the received packet
carries an NSH header, the output port of the primary path is stored
in $reg2$ and the packet is resubmitted to tables ID~1 and ID~3. If a
learning flow exists in table ID~1 for the packet’s output port,
$reg3$ is updated with that port value; otherwise $reg3$ retains the
value zero. In table ID~3, the packet is forwarded through the primary
path if the output port is active; if inactive, the switch selects the
alternative path or drops the packet if none is available.

%%%%%%%%%%%%%%%%%%%%%%%%%%%%%%%%%%%%%%%%%%%%%%%%
\subsection{Link Discovery Protocol}
\label{sec:c-adv}

\noindent Once switches are in the managed state, the controller is aware of the
tree of links connecting it to the rest of the switches. These
connections define the primary path from the controller to each switch
and vice versa. However, any redundant links that are not part of the
primary path remain unknown to the controller. Periplus therefore uses
a link discovery protocol to find additional links that can be used as
alternative paths or to improve the primary path.

\begin{figure}[t]
    \centering

    % Row 1
    \subfloat[$c0$ sends C-Adv.\label{fig:C-Adv1}]
    {
        \includegraphics[width=0.35\linewidth]{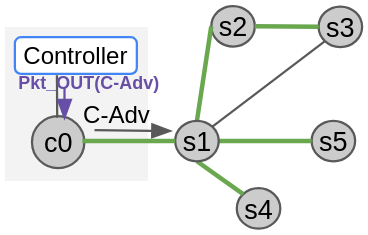}
    }
    \hfill
    \subfloat[$s1$ reports link c0-s1.\label{fig:C-Adv2}]
    {
        \includegraphics[width=0.35\linewidth]{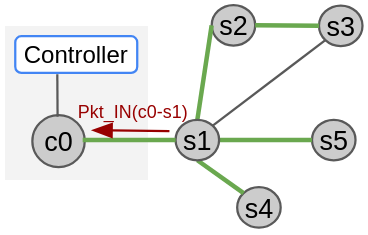}
    }

    %\\[2mm] % espacio entre filas

    % Row 2
    \subfloat[$s1$ sends C-Adv.\label{fig:C-Adv3}]
    {
        \includegraphics[width=0.3\linewidth]{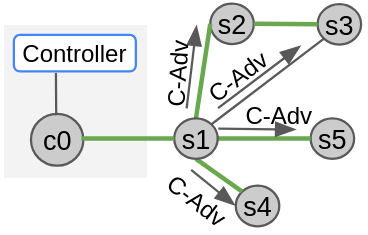}
    }
    \hfill
    \subfloat[$s2$, $s3$, $s4$ and $s5$ report links: $s1-s2$, $s1-s3$, $s1-s4$ and $s1-s5$.\label{fig:C-Adv4}]
    {
        \includegraphics[width=0.3\linewidth]{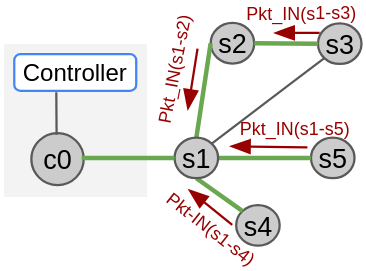}
    }

    %\\[2mm] % espacio entre filas

    % Row 3
    \subfloat[$s2$ and $s3$ send C-Adv.\label{fig:C-Adv5}]
    {
        \includegraphics[width=0.3\linewidth]{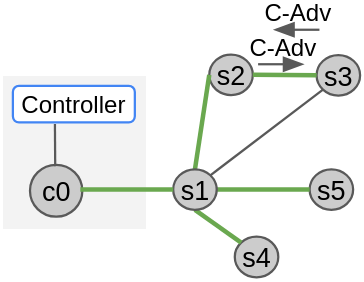}
    }
    \hfill
    \subfloat[$s2$ and $s3$ report links: $s3-s2$, $s2-s3$.\label{fig:C-Adv6}]
    {
        \includegraphics[width=0.3\linewidth]{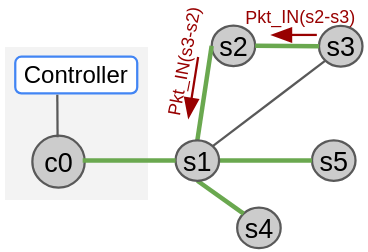}
    }

    \caption{Step-by-step illustration of the C-Adv-based link discovery protocol on a single-controller topology.}
    \label{fig:c-adv-1cs}
\end{figure}

The discovery protocol used by Periplus is based on the periodic
transmission of C-Adv messages by the controller. The controller
issues a Packet-Out to the co-located root switch, which forwards the
message through all its interfaces. Each switch floods the message
only upon first reception, discarding duplicates and thereby enabling
controlled flooding. Duplicate suppression relies on a learning flow
with a hard timeout shorter than the C-Adv period: upon first
reception, the switch installs this flow to mark the current round as
seen; the flow expires before the next round begins, resetting the
per-round state automatically without leaving the learn entry
installed. In
Fig.~\ref{fig:c-adv-1cs}, the green links represent the spanning tree
computed by the controller. The primary path from $s3$ through $s2$ is
longer than the alternative path through $s1$ because the controller
is unaware of the link between $s1$ and $s3$. Fig.~\ref{fig:C-Adv1}
shows the root switch flooding C-Adv.

When a switch receives a C-Adv message, it reports the event to the
controller through a Packet-In (see Fig.~\ref{fig:C-Adv2}). The
Packet-In already determines the receiver-side endpoint of the link:
the OpenFlow connection identifies the receiving switch, and the
\texttt{in\_port} metadata identifies the port on which the C-Adv
arrived. The sender-side endpoint, however, is not visible in standard
Packet-In metadata. To convey it, each managed switch encodes its
own identity and the egress port into header fields of every C-Adv
it forwards; the controller decodes those fields on reception to
identify the sending switch and the egress port. These rewrites are
performed by flow rules installed on the managed switches.

Figs.~\ref{fig:C-Adv3} and~\ref{fig:C-Adv4} illustrate the flooding
performed by switch $s1$ and the subsequent Packet-In messages
generated by switches $s2$, $s3$, $s4$, and $s5$ upon receiving the
C-Adv. Switches $s4$ and $s5$ do not have outgoing links to continue
the flooding process. In contrast, $s2$ and $s3$ both flood the C-Adv
through the interfaces on which it has not yet been received, and each
generates a corresponding Packet-In message, as shown in
Figs.~\ref{fig:C-Adv5} and~\ref{fig:C-Adv6}. This mechanism
effectively prevents broadcast storms of C-Adv messages.

\begin{table}[h]
\caption{Comparison of control-message overhead.}
\label{tab:discMessages}
\scriptsize
\centering
\setlength{\tabcolsep}{3pt}
    \begin{tabular}{|c|c|c|c|}
    \hline
    \textbf{Protocols} & \textbf{\text{Packet-Out}} & \textbf{Link-local messages} & \textbf{\text{Packet-In}}\\
    \hline
        OFDPv1
        & $\sum\limits_{i=1}^{N}\!\texttt{n\_ports}_i$
        & $\sum\limits_{i=1}^{N}\!\texttt{n\_ports}_i$
        & $\sum\limits_{i=1}^{N}\!\texttt{n\_ports}_i$\\
    \hline
    OFDPv2
        & $N$
        & $\sum\limits_{i=1}^{N}\!\texttt{n\_ports}_i$
        & $\sum\limits_{i=1}^{N}\!\texttt{n\_ports}_i$\\
    \hline
    C\text{-}Adv
        & $1$
        & $\sum\limits_{i=1}^{N}\!\texttt{n\_ports}_i\!-\!(N\!-\!1)$
        & $\sum\limits_{i=1}^{N}\!\texttt{n\_ports}_i\!-\!(N\!-\!1)$ \\
    \hline
\end{tabular}
\end{table}

Let $N$ denote the number of switches and $\texttt{n\_ports}_i$ the
number of ports on switch $s_i$. Table~\ref{tab:discMessages}
compares our protocol with OFDPv1 and OFDPv2 on three message
counts: Packet-Out
messages from the controller, link-local messages (LLDP for
OFDPv1/OFDPv2, C-Adv for our protocol), and Packet-In messages from
the switches.

Table~\ref{tab:discMessages} shows that C-Adv reduces Packet-Out
messages to one per discovery round, and reduces both link-local
messages and Packet-In messages by $N-1$ compared to OFDPv1 and
OFDPv2, where $N$ is the number of switches.

\section{Evaluation}
\label{sec:evaluation}

\noindent This section evaluates Periplus along the dimensions established by
its design. We first measure how quickly switches reach the managed
state and how much in-band data this attachment exchanges
(Sections~\ref{sec:switchManagementTime},
\ref{sec:bootstrapTrafficOverhead}), to measure how bootstrap
cost scales with topology diameter and network size. We then
quantify the failure resilience provided by the encoded alternative
paths of the forwarding-graph design, under single link/switch
failures and under cascading failures that exhaust those
alternatives
(Sections~\ref{sec:linkSwitchFailures},
\ref{sec:cascadingFailure}). Finally, we measure the per-switch
flow-table cost of source-hop alternatives
(Section~\ref{sec:flowtableoccupancy}).

The experiments were conducted on an Ubuntu 24.04 system with an AMD
Ryzen 7 4800H CPU (1.4--2.90~GHz) and 64~GB of RAM. Topologies are
emulated with Mininet~2.3~\cite{deOliveira2014mininet}, switches are
OVS~2.15, and Periplus runs on the Ryu~4.32 SDN framework.
Unless otherwise noted, each experiment is repeated twenty times
per topology.
By default, Mininet starts
each OVS instance in the root network namespace; however,
this setup does not allow evaluation of a pure in-band control plane,
where each switch must discover the network before reaching a
controller. To enable such evaluation, our experiments launch each
OVS instance in a separate network namespace and in secure
mode. The separate namespace gives each instance its own isolated
network stack, so that an OVS only sees the veth links to its
physical neighbours, just as a hardware switch would. Secure mode
(OVS's \texttt{fail-mode=secure}) disables the native
learning-bridge fallback: with no controller connection the bridge
installs no rules of its own, behaving as a plain OpenFlow datapath
that drops every packet not explicitly matched by a
controller-installed flow. Together, these settings ensure that a
switch cannot reach a controller until a neighbouring switch is itself
managed and the bootstrap procedure, described in
Section~\ref{sec:bootstrap}, has installed the necessary flows.

\begin{figure}[t!]
\centering
\includegraphics[width=\linewidth]{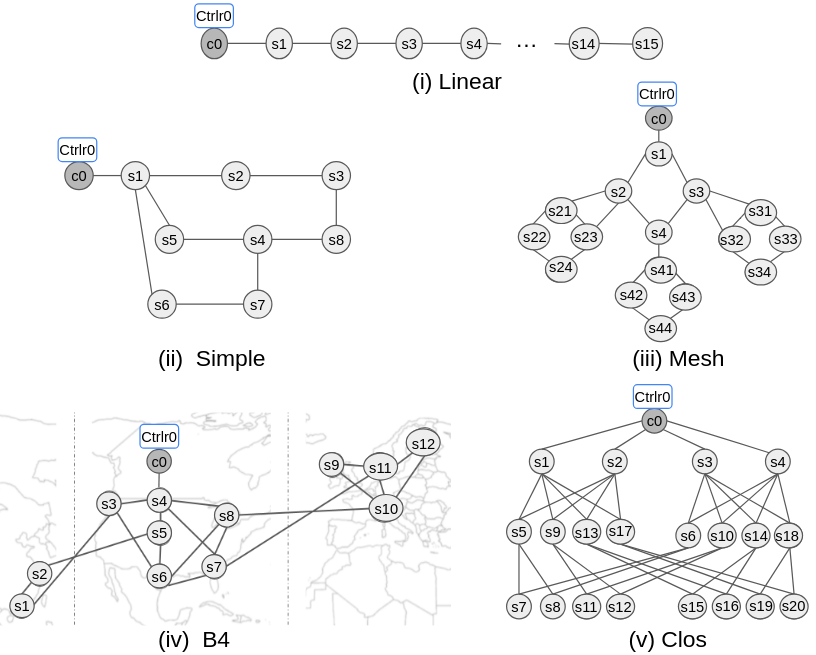}
	\caption{Topologies using one controller: (i) Linear, (ii) Simple, (iii) Mesh, (iv) B4 and (v) Clos.}
\label{fig:oneController}
\end{figure}

Fig.~\ref{fig:oneController} shows the topologies used for the single-controller experiments:
i) Linear (15 switches arranged in a chain),
ii) Simple (8 switches connected through alternative paths to the controller),
iii) Mesh (16 switches forming three interconnected zones),
iv) B4 (Google's 2011 WAN deployment~\cite{jain2013b4}), and
v) Clos (20 switches arranged in a three-stage topology).

\subsection{Switch management time}
\label{sec:switchManagementTime}

\noindent A direct implication of the wave-by-wave attachment of
Section~\ref{sec:bootstrap} is that the incremental anchor-relayed
procedure scales with network diameter rather than with the number
of switches: a switch can begin its attachment only once a
neighbouring switch is itself managed, so the critical path
through the topology determines the minimum bootstrap time. 
We verify this empirically with two complementary
metrics: the \emph{Periplus per-switch bootstrap time}, which sums the
per-hop attachment delays along the critical path and isolates the
protocol cost, and the \emph{wall-clock bootstrap time}, which
measures the interval from the first TCP SYN received at any
controller to the moment the last switch is marked managed across all
controller logs.

We use the five topologies of Fig.~\ref{fig:oneController}, each
with a single controller. The
experiments launch the controller before the regular switches, and
start the regular switches in parallel, so that each switch
discovers its upstream port from the first C-Adv or ARP reply it
receives and immediately begins its bootstrap.

\subsubsection{Periplus per-switch bootstrap time}
\begin{figure}[t!]
    \centering
	\includegraphics[width=0.78\linewidth]{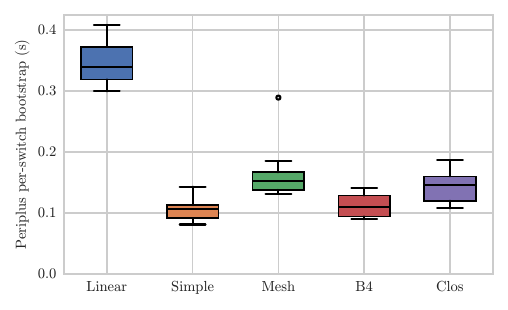}
\caption{Periplus per-switch bootstrap time for the five
  topologies.}
\label{fig:periplusAll}
\end{figure}

\noindent Fig.~\ref{fig:periplusAll} shows per-switch bootstrap time
(per-rep maximum across switches). Medians range from
$0.11$~s (Simple) to $0.34$~s (Linear, diameter~15).

\begin{figure}[t!]
    \centering
	\includegraphics[width=0.78\linewidth]{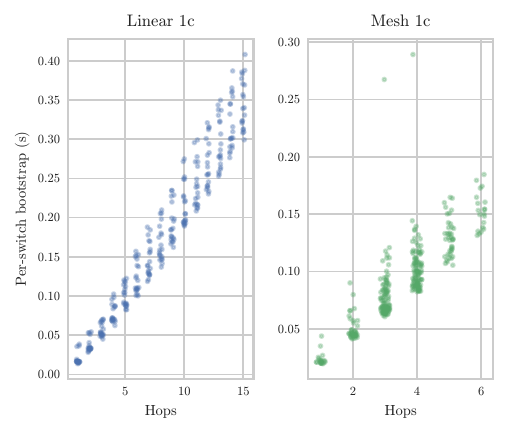}
\caption{Per-switch bootstrap time as a function of hop distance
  from the controller, for Linear and Mesh. Each dot is one
  (switch, repetition) sample with small horizontal jitter.}
\label{fig:timeVsDistance}
\end{figure}

Fig.~\ref{fig:timeVsDistance} resolves the per-switch cost by hop
distance. Linear (left) grows linearly at $\approx 23$~ms per hop,
consistent with the wave-by-wave attachment of
Section~\ref{sec:bootstrap}. Mesh (right) shows wider spread at
each distance because equal-cost paths compete for the anchor
role.

\subsubsection{Wall-clock bootstrap time}

\begin{figure}[h]
    \centering
	\includegraphics[width=0.78\linewidth]{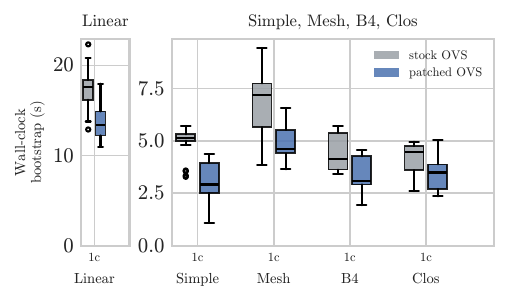}
\caption{Wall-clock bootstrap time for the five topologies.
  Linear is given its own sub-panel because its stock
  value (median $\approx 18$~s) would otherwise compress every
  other topology; Simple, Mesh, B4 and Clos share the right
  sub-panel. Grey boxes are stock OVS; blue boxes are patched OVS.
  Outlier reps beyond twice the median are excluded.}
\label{fig:wallclockComparison}
\end{figure}

\noindent Fig.~\ref{fig:wallclockComparison} reports wall-clock bootstrap.
Grey boxes are stock OVS, blue boxes are a build with a small patch
to OVS's \texttt{rconn} reconnect module that removes hardcoded
multi-second floors on the connection-retry and inactivity-probe
intervals.\footnote{The same patch also corrects how OVS interprets
activity on an established OpenFlow session, a behaviour that is
relevant after a controller failure but not measurable in the
bootstrap scenario reported here.}
The patch does not change the Periplus protocol; it only shortens
the time a switch waits before retrying the connection to its
upstream while the upstream is not yet managed. Stock takes
$1.3$--$1.8\times$ as long as patched on every topology, the gap
dominated by rconn floors accumulated along the cascade of
attachments.

\subsection{Bootstrap overhead}
\label{sec:bootstrapTrafficOverhead}

\noindent Beyond how quickly a switch can be brought under control, an operator
also cares about how much in-band data that attachment exchanges. The per-switch bootstrap cost is expected to be dominated by the
OpenFlow handshake plus a small set of flow installations whose
size depends on \emph{local} topology structure (specifically, on
how many loop-free alternatives each hop of the primary path has)
and \emph{not} on the total number of destinations in the
network. To measure this, we record the volume of Ethernet-level
data exchanged between the controller and each newly-attached
switch during its bootstrap.

We capture all Ethernet frames exchanged between the controller and
each newly attached switch, from the moment that switch's TCP SYN
arrives at the controller's co-located switch until the controller
marks it as managed. The controller's co-located switch is itself
excluded because it bootstraps locally rather than through the in-band
procedure. Four topologies are reported: Linear, Simple and Mesh from
Fig.~\ref{fig:oneController}, plus a degenerate \emph{Single}
topology with one managed switch directly attached to the controller's
co-located switch, included as a baseline to show the OpenFlow-handshake
floor with no graph-embedding flows installed.

\begin{figure}[h]
    \centering
	\includegraphics[width=0.6\linewidth]{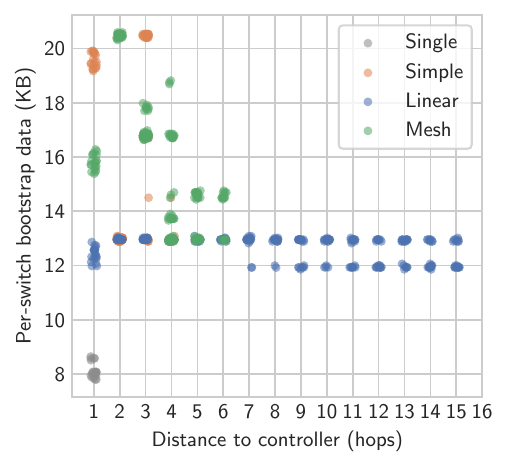}
\caption{Per-switch bootstrap data volume versus the switch's
  hop distance to the controller, for four topologies (Single,
  Simple, Linear and Mesh). One point per
  (switch, repetition) sample, with small horizontal jitter so
  coincident points are visible.}
\label{fig:bootstrapTraffic}
\end{figure}

Fig.~\ref{fig:bootstrapTraffic} reveals two regimes. Linear is
flat at $\approx$13~KB across hop distances 1--15: the per-switch
cost is dominated by the $\approx$13~KB OpenFlow-handshake floor
and does not grow with hop distance. Single sits below at
$\approx$8~KB.
Mesh and Simple spread wider because some hops carry multiple
loop-free alternatives. The per-switch cost is thus
$O(\sum_i K_i)$, the total number of encoded alternatives along the
path, and independent of the total number of destinations,
consistent with the per-switch state property of
Section~\ref{sec:forwarding-graphs}.

\subsection{Link/Switch failures}
\label{sec:linkSwitchFailures}

\noindent How quickly can the network absorb a link failure when each packet
already carries an encoded alternative path? Periplus's
forwarding-graph design (Section~\ref{sec:forwarding-graphs})
targets sub-50~ms recovery without controller involvement: a
switch detecting a failed output port falls over to the encoded
alternative immediately, before any control-plane reaction is
possible. We test this target by quantifying the throughput gap
caused by a single link failure on a long-lived flow.

\begin{figure}[h]
    \centering
	\includegraphics[width=0.5\linewidth]{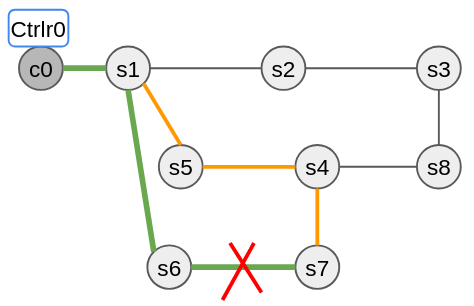}
\caption{Simple topology (Section~\ref{sec:linkSwitchFailures}) with the link s6--s7 marked as the failure point. The primary path s7$\rightarrow$s6$\rightarrow$s1$\rightarrow$c0 (green) and the alternative path through $s4$ and $s5$ (orange) are highlighted.}
\label{fig:failure-s6-s7}
\end{figure}

\begin{figure}[b!]
    \centering
	\includegraphics[width=0.6\linewidth]{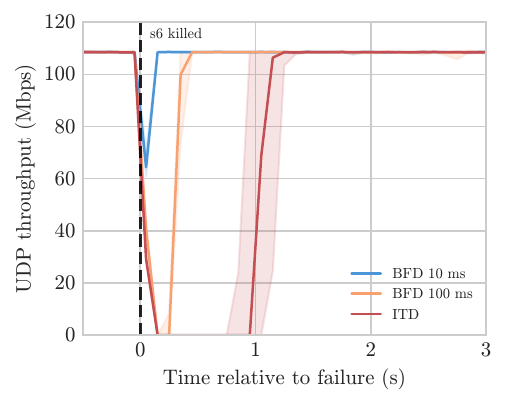}
\caption{UDP throughput around a link failure for three
  failure-detection mechanisms: BFD at 10~ms, BFD at 100~ms, and ITD.
  Median trace with the 25--75 percentile band. Throughput is
  measured at the Ethernet level, hence the plateau above the
  100~Mbps offered rate.}
\label{fig:nodeLinkFailure}
\end{figure}

We use the Simple topology of Fig.~\ref{fig:failure-s6-s7}, with the
primary path s7$\rightarrow$s6$\rightarrow$s1$\rightarrow$c0 shown in
green and the alternative path through s4 and s5 shown in orange. A
UDP iperf~\cite{iperf2} flow at a 100~Mbps offered rate from s7 to c0
is interrupted by killing switch s6 (vertical dashed line in
Fig.~\ref{fig:nodeLinkFailure}, taken as $t=0$). We compare the three
failure-detection mechanisms supported by Periplus
(Section~\ref{sec:monitoringports}): BFD at 10~ms, BFD at 100~ms, and
ITD with a 2~s learning-flow timeout.

Fig.~\ref{fig:nodeLinkFailure} shows the result. BFD-$10$~ms
gives a $\approx 50$~ms median gap, barely visible at this
scale; BFD-$100$~ms widens to $\approx 350$~ms and ITD to
$\approx 1$~s, proportional to the detection delay. A minority
of BFD-$10$~ms repetitions ($\approx 25$\%) show longer tails
(median $\approx 1$~s among that subset, up to $\approx 2.5$~s):
the encoded alternative is available immediately, so we attribute
the slow resume to a host/OVS-level artifact rather than a Periplus
recovery cost. The figure's median trace reflects the majority. In all three
cases the alternative path immediately resumes the offered
rate with no post-failover penalty: recovery is driven by the
encoded alternative carried by the in-flight packets, not by
the controller. The controller later installs replacement
primary paths via C-Adv, in the background while traffic
already flows along the embedded alternative.

\subsection{Cascading failure recovery}
\label{sec:cascadingFailure}

\noindent We evaluate the response of Periplus to a simultaneous failure
that removes both the primary source-hop port at switch s4 and
the first port of its first encoded alternative. The Simple
topology of Fig.~\ref{fig:oneController} provides s4 with three
node-disjoint paths to the controller c0:
\begin{itemize}
\item \textbf{Primary} $P$:
  s4$\rightarrow$s5$\rightarrow$s1$\rightarrow$s0
\item $A_1^{(1)}$:
  s4$\rightarrow$s7$\rightarrow$s6$\rightarrow$s1$\rightarrow$s0
\item $A_1^{(2)}$:
  s4$\rightarrow$s8$\rightarrow$s3$\rightarrow$s2$\rightarrow$s1$\rightarrow$s0
\end{itemize}
By symmetry, c0 reaches s4 along the same three paths in reverse.

\begin{figure}[t!]
    \centering
	\includegraphics[width=0.6\linewidth]{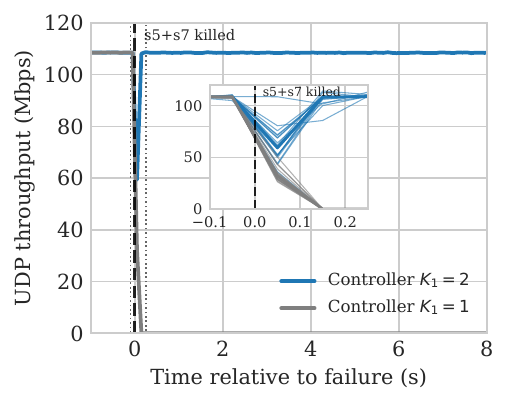}
\caption{Cascading failure: UDP throughput at 100\,Mbps from s4 to
  c0 around the simultaneous kill of s5 and s7, with BFD\,10\,ms
  detection plus ITD. The controller started with $K_1 = 2$
  source-hop alternatives (blue) recovers in the data plane via
  $A_1^{(2)}$; the controller started with $K_1 = 1$ (gray) has
  no preinstalled alternative left and waits for control-plane
  re-routing. Inset zooms on the $K_1 = 2$ recovery interval.
  Median trace; the dashed vertical line marks
  the failure.}
\label{fig:cascadingFailure}
\end{figure}

We kill s5 and s7 simultaneously, removing the source-hop ports
used by $P$ and $A_1^{(1)}$ at s4 and the corresponding return
paths at c0. Failure detection uses BFD at a 10\,ms detection
interval combined with ITD running on all switches. A long-lived
UDP iperf flow at a fixed offered rate of 100\,Mbps runs from s4
to c0.
Fig.~\ref{fig:cascadingFailure} overlays two controller startup
configurations.

With the controller started with $K_1 = 2$ (blue), the
install-time encoding places $P$, $A_1^{(1)}$ and $A_1^{(2)}$ in
s4's packet headers toward c0, and the three symmetric reverse
paths in c0's flow rules toward s4. After BFD detects that both
s5 and s7 are down, s4 falls through to $A_1^{(2)}$
(s4$\rightarrow$s8$\rightarrow$s3$\rightarrow$s2$\rightarrow$s1$\rightarrow$s0)
following Algorithm~\ref{alg:periplus_forwarding}, entirely in
the data plane. The reply traffic from c0 follows the symmetric
reverse path. Median recovery is ${\sim}50$\,ms across 20/20
repetitions, as the inset of Fig.~\ref{fig:cascadingFailure}
shows.

The $K_1=2$ recovery is in the same order of magnitude as the
single-link case of Section~\ref{sec:linkSwitchFailures} at the
same BFD interval. The mechanism itself adds no latency beyond
detection: each $A_1^{(k)}$ candidate is matched in a single
OpenFlow rule lookup.

With the controller started with $K_1 = 1$ (gray), the encoding
places only $P$ and $A_1^{(1)}$ in s4's headers and the symmetric
two paths in c0's rules. With both encoded source-hop ports dead,
the data plane has no preinstalled alternative and recovery falls
back to the control plane, where C-Adv must propagate the failure
to c0 before new routes are installed. The control-plane fall-back
is slower than the data-plane reaction by orders of magnitude and,
in this experiment, exhibits variability that depends on how the
current recovery path handles multi-hop failures; characterizing
this variability is left to future work.

Encoding $K_1 \geq 2$ source-hop alternatives therefore converts
a control-plane round-trip into a data-plane reaction whose
latency is bounded by the failure-detection mechanism, a speedup
of orders of magnitude over the control-plane fall-back.

\subsection{Flow-table cost of source-hop alternatives}
\label{sec:flowtableoccupancy}

\noindent The $K$ source-hop alternatives mechanism of
Section~\ref{sec:forwarding-graphs} extends the source-hop
forwarding chain by one alt-failover rule per source-hop
alternative beyond the primary. We quantify the resulting
per-switch overhead in the Simple topology of
Section~\ref{sec:cascadingFailure}, sweeping
the install-time policy between $K_1 = 1$ (no alternative
preinstalled) and $K_1 = 2$ (one alternative preinstalled).

\begin{table}[h]
\centering
\caption{Per-switch flow-table occupancy, in number of installed
  flow rules, in the Simple topology under $K_1 = 1$ and
  $K_1 = 2$. Median over five repetitions; per-switch variance was
  zero across reps.}
\label{tab:flowTableOccupancy}
\small
\begin{tabular}{|l|r|r|r|}
\hline
\textbf{Switch} & \textbf{$K_1 = 1$} & \textbf{$K_1 = 2$} & \textbf{$\Delta$} \\
\hline
c0 & 29 & 29 & 0 \\
s1 & 61 & 61 & 0 \\
s2 & 42 & 42 & 0 \\
s3 & 52 & 52 & 0 \\
\textbf{s4} & \textbf{62} & \textbf{65} & \textbf{+3} \\
s5\,--\,s8 & 42 & 42 & 0 \\
\hline
\end{tabular}
\end{table}

Table~\ref{tab:flowTableOccupancy} confirms the locality property of
the design. Only s4 has two source-hop alternatives to c0; under
$K_1 = 2$ the controller installs three additional rules at s4
(the $K_1 = 2$ source-primary chain and its alt-failover entries).
The other eight switches have a single source-hop alternative each
toward c0 and their occupancy is unchanged. Network-wide, the
increment is 3 rules over a $\approx$378-rule baseline.

The cost grows by one alt-failover chain per (switch, destination)
pair with $K \geq 2$ and is bounded by the topology's path
diversity, not by network size. This complements the per-switch
state property already evidenced by bootstrap data volume in
Section~\ref{sec:bootstrapTrafficOverhead}: the same locality that
bounds bootstrap-time cost also bounds the steady-state flow-table
footprint of source-hop alternatives.

\section{Conclusions and Future Work}
\label{sec:conclusions}

\noindent This paper presents Periplus, an SDN in-band control plane that
jointly addresses four challenges of in-band SDN: bootstrapping,
routing, fast failure recovery, and multi-controller coordination.
The first three are developed in this paper; multi-controller
coordination will be addressed in a separate paper. For bootstrapping,
Periplus uses Proxy ARP and Anycast to discover the controller
without flooding ARP or TCP traffic through the network; upon
detecting a new switch, the controller installs flow rules in only
two switches, rather than updating all switches on the path to the
controller as in prior approaches. For routing, Periplus embeds forwarding graphs into packet
headers to enable source-based routing with per-hop alternative
paths, requiring no modifications to OVS. Each switch stores only the path to its own controller,
while the root switch alone maintains the paths to all switches in its
domain, reducing flow-table occupancy at the cost of encoding path
information in each control packet.

For failure recovery, Periplus detects failures through BFD and
ITD and reroutes traffic immediately using the
alternative paths encoded in packet headers, without contacting the
controller. This achieves sub-50\,ms reaction times and maintains
controller-to-switch connectivity after a failure as long as an
alternative path exists.

The Mininet evaluation across topologies of varying size and
diameter validates each of these design choices. Bootstrap scales
with topology diameter rather than network size: per-switch
medians stay below 0.34\,s. Per-switch bootstrap traffic is
dominated by the $\approx$13\,KB OpenFlow handshake; the
graph-encoding cost adds a few hundred bytes per loop-free
alternative along the path, independent of the total number of
destinations. Failure recovery with BFD at a 10\,ms detection
interval achieves sub-50\,ms switchover with no throughput
degradation once the alternative path is engaged. Adding a
second source-hop alternative ($K_1 = 2$) converts cascading-
failure recovery from a control-plane round-trip into a data-plane
reaction bounded by the failure-detection mechanism, at a per-
switch flow-table cost local to the switches with two source-hop
alternatives (three additional rules in the Simple topology).

One direction for future work targets the bandwidth consumed by
Periplus itself: each forwarding-graph path is currently carried
in its own 128-bit NSH header, which is often only partially used.
A more compact, variable-length encoding that packs short
alternative paths together would reduce both the per-packet
overhead and the size of the flow-mod messages installed at
bootstrap.

Finally, extending Periplus to wireless and hybrid wired-wireless
networks requires a virtual port abstraction to handle ephemeral,
MAC-layer-identified links, with corresponding changes to
bootstrapping, forwarding graph encoding, and failure detection; a
prototype for hybrid wired-wireless topologies is currently under
development.

\bibliographystyle{plain}
\bibliography{refs}

\end{document}